\documentclass[conference,transmag]{IEEEtran}
\IEEEoverridecommandlockouts

\usepackage{cite}
\usepackage[utf8]{inputenc}
\usepackage{amsmath,amssymb,amsfonts}
\usepackage{algorithmic}
\usepackage{graphicx}
\usepackage{textcomp}
\usepackage{xcolor}
\usepackage{verbatim} 
\usepackage{paralist}
\usepackage{hyperref}
\usepackage{tabularx}
\hypersetup{
  hidelinks, 
}

\newcommand{\ab}[1]{{\color{green} AB: #1}}

\begin{document}

\title{Evaluation of Cognitive Architectures for Cyber-Physical Production Systems
}
\author{\IEEEauthorblockN{Andreas Bunte\IEEEauthorrefmark{1}$^*$\thanks{$^*$Authors contributed equally to this paper.}, Andreas Fischbach\IEEEauthorrefmark{2}$^*$, Jan Strohschein\IEEEauthorrefmark{3}$^*$,\\
Thomas Bartz-Beielstein\IEEEauthorrefmark{2}, Heide Faeskorn-Woyke\IEEEauthorrefmark{3} and Oliver Niggemann\IEEEauthorrefmark{1}} \\
\IEEEauthorblockA{\IEEEauthorrefmark{1}OWL University of Applied Sciences, Institute Industrial IT, Lemgo, Germany\\
Email: andreas.bunte@hs-owl.de, oliver.niggemann@hs-owl.de}
\IEEEauthorblockA{\IEEEauthorrefmark{2}TH Köln, Institute for Data Science, Engineering, and Analytics, Gummersbach, Germany\\
Email: andreas.fischbach@th-koeln.de, thomas.bartz-beielstein@th-koeln.de}
\IEEEauthorblockA{\IEEEauthorrefmark{3}TH Köln, Institute of Computer Science, Gummersbach, Germany\\
Email: jan.strohschein@th-koeln.de, heide.faeskorn-woyke@th-koeln.de}
}
\newpage
\begin{titlepage}
\begin{Large}
\vspace*{4cm}
\textcopyright 2019 IEEE.  Personal use of this material is permitted.  Permission from IEEE must be obtained for all other uses, in any current or future media, including reprinting/republishing this material for advertising or promotional purposes, creating new collective works, for resale or redistribution to servers or lists, or reuse of any copyrighted component of this work in other works.
\end{Large}
\end{titlepage}

\maketitle
\begin{abstract}
Cyber-physical production systems (CPPS) integrate physical and computational resources due to increasingly available sensors and processing power. 
This enables the usage of data, to create additional benefit, such as condition monitoring or optimization. 
These capabilities can lead to cognition, such that the system is able to adapt independently to changing circumstances by learning from additional sensors information. 
Developing a reference architecture for the design of CPPS and standardization of machines and software interfaces is crucial to enable compatibility of data usage between different machine models and vendors.
This paper analysis existing reference architecture regarding their cognitive abilities, based on requirements that are derived from three different use cases.
The results from the evaluation of the reference architectures, which include two instances that stem from the field of cognitive science, reveal a gap in the applicability of the architectures regarding the generalizability and the level of abstraction. While reference architectures from the field of automation are suitable to address use case specific requirements, and do not address the general requirements, especially w.r.t. adaptability, the examples from the field of cognitive science are well usable to reach a high level of adaption and cognition. It is desirable to merge advantages of both classes of architectures to address challenges in the field of CPPS in Industrie 4.0.
\end{abstract}
\begin{IEEEkeywords}
Reference Architecture, Cognition, Industrie 4.0
\end{IEEEkeywords}

\section{Introduction}
"Cyber-Physical Systems (CPS) are integrations of computation with physical processes"\cite{Lee2008}. If a CPS is used in a production setting it is called a cyber-physical production system (CPPS)\cite{Bangemann2015}. CPPS integrate physical and computational resources deeper than it was possible before because of increasingly available sensors and processing power, which is a part of "Industrie 4.0" (I4.0). I4.0 enables new insights into processes and the life-cycle of assets used in production. Within companies the shop floor and the office floor can work together in a more efficient manner. For example new high priority orders can be added to the production, which dynamically updates the plans to fulfill all orders in time. Additionally integration of information systems could help to include suppliers and subcontractors better into the business processes of a company. The new level of integration creates the need to unify the efforts of all industries related to I4.0, e.g., mechanical engineering, electrical engineering and IT. Whenever different fields work together it is important to create commonly accepted definitions and standards to make collaboration easy and efficient. Therefore developing a reference architecture for the design of CPPS and standardization of machine and software interfaces is crucial to enable compatibility between different machine models and vendors (see also \cite{Bitkom:2015}). In contrast to this only 14\% of executives believe that their organization is prepared for I4.0 and able to realise the additional potential\cite{Deloitte2018}. Executives and workers alike need assistance to implement necessary changes in order to benefit from further developments. This assistance can be given by a reference architecture, which defines and specifies common features and provides a guideline for software engineers to design a CPPS. They ensure compatibility of components and efficient software development for CPPS. To fully utilize the potential of a CPPS in the context of I4.0 the systems need cognitive abilities. A cognitive CPPS is able to learn from the additional information collected from sensors and adjust independently to changing circumstances. The system contains abilities like learning, planning and decision making to adapt to new situations such as production planning based on a new order, failure of a module but also optimizing production and detecting anomalies before any component malfunctions. 
So ultimately a cognitive reference architecture for CPPS should be constructed.

A set of different use cases has been analyzed in order to collect real-world demands towards a reference architecture. Comparing the different use cases reveals an intersection of requirements that can be generalized into a reference architecture. 
Researchers and organizations proposed several reference architectures to design CPPS, e.g., Reference Architecture Model Industrie 4.0 (RAMI4.0), Industrial Internet Reference Architecture (IIRA) and 5C. Additionally two architectures, Soar and Adaptive-Control-of-Thought-Rational (ACT-R), that stem from the field of cognitive sciences are presented as cognitive abilities are also required. 
This paper will examine those architectures regarding chosen CPPS use cases and contribute in three ways:
\begin{enumerate}
	\item [(C1)] Identify requirements of cognitive architectures
	\item [(C2)] Evaluation of proposed reference architectures
	\item [(C3)] Identify gaps in proposed reference architectures
\end{enumerate}
To accomplish this, the next section introduces use cases and highlights requirements towards a reference architecture. Subsequently the selection of proposed reference architectures is presented and evaluated against the derived requirements, in Section\,\ref{sec:evaluation}. The resulting gap will be discussed in the closing section.
\section{Use Cases and Requirements} \label{sec:use-case}

The following section introduces use cases which are used to derive requirements towards a reference architecture. Those requirements are subsequently condensed in Table\,\ref{tab:OverallRequ}. An early utilization of proposed reference architectures will highlight benefits but also areas that still need improvements.

\subsection{Diagnosis of Modular CPPS}
The versatile production system (VPS) in the SmartFactoryOWL is a demonstration plant of a CPPS. 
The VPS processes corn in different modules and finally produces popcorn out of it.
It consists of the modules delivery, storage, quality control, dosing and production. 
Modules have compatible interfaces, which allow different hardware configurations to enable adaptive production.
It is also possible the extend the system with new modules, which might be unknown during the engineering phase of the initial modules.
Therefore it is a self-diagnosis use case which implements a diagnosis system, such as \cite{Bunte:2019}, to detect anomalies independently of the current configuration, determine the root-cause and stop the faulty modules to prevent additional damage.
All this, should be performed by artificial intelligence algorithms.
This is an example of value-based services, sub scenario \textit{Condition Monitoring Services} from the Plattform Industrie 4.0\,\cite{VBS:2017}, but the self-reaction of the CPPS goes beyond that use case. 
Due to the versatile combination possibilities, no diagnosis system is implemented in the VPS, since approaches known today cannot handle these versatile systems efficiently.
Therefore, data driven approaches can be an efficient method for anomaly detection, as it is done in\,\cite{Maier:2015}.
These methods provide a huge potential, but their implementation is difficult, because there are no common interfaces and thus it has to be adapted for every specific application with a certain effort.
The architecture must address the following requirements:
\begin{compactenum}
  \item[(R-A.1)] Data acquisition via Open Platform Communications Unified Architecture (OPC UA) server and Modbus. 
  \item[(R-A.2)] Store data to collect all lifecycle information, such as process data and models of the CPPS.
  \item[(R-A.3)] Perform a preprocessing to handle missing values, normalize data or adapt the format. 
  \item[(R-A.4)] Learn diagnosis models that enable diagnosis with few learning data sets for online and offline learning.
  \item[(R-A.5)] Provide a diagnosis algorithm that performs diagnosis in less that 100 ms. 
  \item[(R-A.6)] Decision making whether or not a response is required and choosing an action.
  \item[(R-A.7)] Access the controller to perform the chosen action and thus prevent further damage.
\end{compactenum}

\subsection{Energy Efficiency Optimization in Bakeries}
Optimizing the usage of energy is an important task especially in industries with a high energy consumption. 
Consider bakeries that consists of several chain stores of different types, i.e., sale only, production only and stores combining both, production and sale. 
Production devices, especially ovens, are the major energy consumption devices. 
Ovens contain several herds, which can be controlled individually.
Planning an optimal baking procedure with different products at different temperatures is a crucial task w.r.t. energy efficiency, as idle times for the herds and unnecessary cool-down times are to avoid.
Companies' energy costs are not only calculated by the consumption, also the maximum used power has to be payed. Thus for companies it is beneficial to divide the energy consumption over the day, instead of creating large peaks by, e.g., turning on all devices at start of business.
Thus a smart start up schedule provides a significant contribution in cost optimization.
Furthermore the stores typically also have a large divergence regarding the numbers of arriving customers at different times during day, and thus necessary stock of products. 
Additionally to the monitoring of energy consumption, the predicted amount of products to sell and the available stock of products should be regarded. 
A smart integration of the inventory and sale system enables the computation of accurate decentralized models of product sales for a given time of each day. 
Enabling information exchange between different models can lead to a higher level of adaptability. 
The architecture must address the following requirements:
\begin{compactenum}
  \item[(R-B.1)] Highly distributed and heterogeneous systems. 
  \item[(R-B.2)] Store data, such that suitable historical data can be used, e.g., for optimization. 
  \item[(R-B.3)] Preprocess data, e.g., treat missing data accordingly.
  \item[(R-B.4)] Learn and update models of energy consumption. 
  \item[(R-B.5)] Provide a simulation to enable energy efficiency optimization and peak prevention.
  \item[(R-B.6)] Provide guidance to employees by implementing an appropriate human machine interface (HMI).
\end{compactenum}

\subsection{Process Control of Concrete Spreading Machines}
A concrete spreading machine produces pre-cast concrete components. 
The mold consists of a steel pallet, casing and additional reinforcements. 
The machine pours concrete on the steel pallet.  
Casing and reinforcements mounted on the steel pallet determine the shape and properties of a component. Controlling parameters for the process have to be set manually, which is quite difficult and needs a lot of experience since some parameters are hard to assess, e.g., the consistency of concrete from an earlier cast mixed with fresh concrete. Additionally changing parameters will not lead to an immediate reaction as the concrete moves slowly. The goal of this use case is to learn a model that is able to control the process in order to optimize two conflicting goals: minimizing the production time while maximizing the quality of the resulting product. 
In order to implement a system that learns from process data to control and optimize the production output the following requirements towards an architecture emerge:
\begin{compactenum}
  \item[(R-C.1)] Ingest sensor data and perform preprocessing to extract standardized process information.
  \item[(R-C.2)] Store relevant information of the production process.
  \item[(R-C.3)] Learn models from historical as well as current sensor data to control the process and consider conflicting goals.
  \item[(R-C.4)] Decision making in real-time or near-real-time to be useful in a real-world production process.
  \item[(R-C.5)] Verify results in a simulation before allowing the algorithm to control an actual concrete spreading machine. 
  \item[(R-C.6)] Communicate with the machine in a standardized format for compatibility with different models/vendors.
\end{compactenum}

\subsection{General Requirements}
The aim of the cognitive reference architecture is to cover versatile use cases within one architecture. 
Therefore implementations can be easily adapted to different use cases which reduces engineering costs.
In a specific use case, the adaption might be the selection of a proper algorithm, based on the user specification.
The architecture must enable the users to implement decision making and learning which decisions are promising for which application.
To achieve this, there are some additional requirements that cannot be derived from the single use cases, so there are some overall requirements:

\begin{compactenum}
	\item[(R-D.1)] Receive declarative goals from the user.
	\item[(R-D.2)] Specified interfaces are well defined.
	\item[(R-D.3)] Strategies to select a suitable algorithm.
	\item[(R-D.4)] The system learns from experiences.
	\item[(R-D.5)] Thorough knowledge representation. 
\end{compactenum}

The user should be able to define goals easily. If the system accepts declarative goals (R-D.1) it is possible to define a goal as plain as "optimize energy consumption". Additionally the user can specify additional constraints, such as limitation of the response time. Furthermore the reference architecture should possess well defined interfaces (R-D.2) with a thorough description which information should be transferred to ensure modular expandability. Re-usability and customization benefit from such a modular structure and this will also enable concepts such as software as a service. Additionally exchange and purchase of services between vendors are possible. The reference architecture needs a strategy to select an algorithm (R-D.3) that is able to produce a feasible result under the current conditions, such as volume of data, the required response time or the type of problem to solve. A cognitive architecture reflects upon the decision making and is able to learn from past experiences (R-D.4) to increase processing efficiency and outcome over time. Essentially what was the right tool for a given job. To aid this decision making and to model/store additionally learned knowledge it is required to implement a suitable knowledge representation (R-D.5) that represents information about machines and processes, newly learned rules and domain knowledge from experts. 
\section{Reference Architectures}
There are two different classes of architectures that are related to this work, namely reference architectures from the field of automation and cognitive architectures with a background of cognitive sciences.
We do not consider models that are used to analyse the current state of a production, e.g., \textit{'Reifegradmodell Industrie 4.0'}~\cite{Jodlbauer:2016}. 
The reference architectures in automation present a generic structure for a class of architectures that should help to design automation systems.
We chose some well known architectures, namely the RAMI4.0, IIRA, and the 5C architecture.
In comparison to that, traditional cognitive architectures from the field of cognitive sciences have been built to understand human cognition.
Such a cognitive architecture is defined in\,\cite{Lehman:1996} as \textit{'... is a theory of the fixed mechanism and structures that underlie human cognition.'}
We introduce the Soar and the ACT-R architectures, from this field. In automation, the word cognitive is used on a higher level than in cognitive sciences.A cognitive architecture in automation adapts the process independently to new situations, such as creating an adjusted production plan, if a module breaks.
This does not contradict the definition above, as it may not require to understand human cognition to solve a high-level goal in automation.

\subsection{Reference Architecture Model Industrie 4.0 (RAMI4.0)}
\begin{figure}[b]
	\centering
	\includegraphics[width=0.4\textwidth]{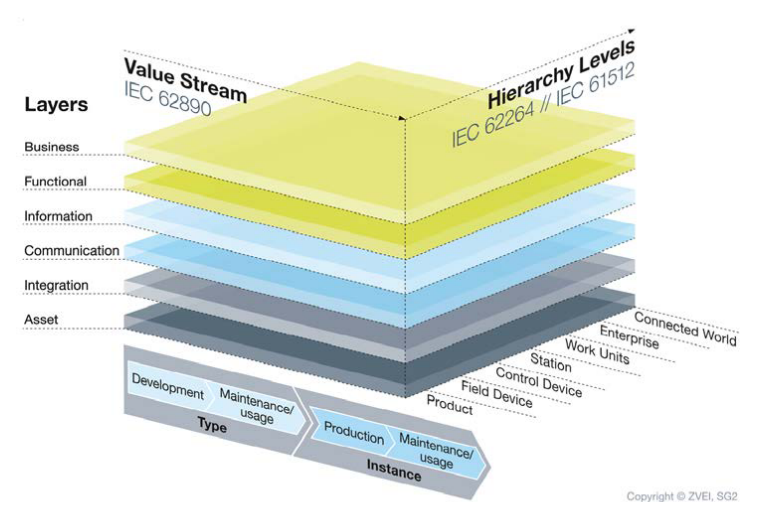}
	\caption{Reference Architecture Model Industrie 4.0 (RAMI4.0).\,\cite{VDI:2015}}
	\label{fig:RAMI4.0}
\end{figure}

In 2013 the three German associations Bitkom, VDMA and ZVEI founded the "Plattform Industrie 4.0" to merge the different interests and requirements of electrical engineering, mechanical engineering and information technology towards a joint model for Industrie 4.0. The resulting reference architecture model, shown in Fig.\,\ref{fig:RAMI4.0}, was published in 2015 \cite{Bitkom:2015} \cite{VDI:2015}.
The model consists of three dimensions to depict the essential aspects: 
\begin{itemize}
	\item Layers: The six layers are based on the Smart Grid Architecture Model \cite{SGAM:2012} and have been customized to fit I4.0 requirements. The layers depict the digital representation of an asset, for example a machine. The digital representation includes the asset specification but also a description of functional and communication behaviour. The complexity of the layers increases up to the \textit{Functional} and \textit{Business} layer. 
	Those provide a runtime and orchestrate the services that support the business processes. The layers are loosely coupled and events and data are meant to be exchanged within a layer or between neighboring layers\cite{VDI:2015}. 
	
	\item Life Cycle \& Value Stream: The horizontal axis represents the life cycle of plants and products in the I4.0 environment on the basis of IEC 62890 and distinguishes types and instances.\cite{VDI:2015}. 

	\item Hierarchy Levels: This dimension classifies assets regarding their I4.0 functionality and responsibilities within a production plant. The classification is based on IEC 62264 and IEC 61512 and extends the automatization pyramid, introducing the additional levels \textit{product} and \textit{connected world} to fulfill the new requirements \cite{Bitkom:2015}. 
\end{itemize}

Therefore RAMI4.0 can be used to classify machines and components of a CPPS and provides a standardized approach for the design of I4.0 machines. This intention was reaffirmed when in the same publication also the first related model was presented. The "Industrie 4.0 Component" contains communication functionality and an administrative shell \cite{VDI:2015}\cite{ZVEI:2016}. Such an I4.0 component could be an entire production system, a single machine or even just a module of a machine. Current assets can be upgraded to I4.0 components as the additional functionality may be provided by an external system. 

\subsection{Industrial Internet Reference Architecture}
The Industrial Internet Consortium (IIC) introduced its Industrial Internet Reference Architecture (IIRA) in 2015 and updated to version 1.8 in 2017 as a standards-based architectural template and methodology to enable Industrial Internet of Things (IIoT) System architects to design their systems based on a common framework and concepts~\cite{Lin2017}. 
\begin{figure}[htb]
  \centering
  \includegraphics[width=0.4\textwidth]{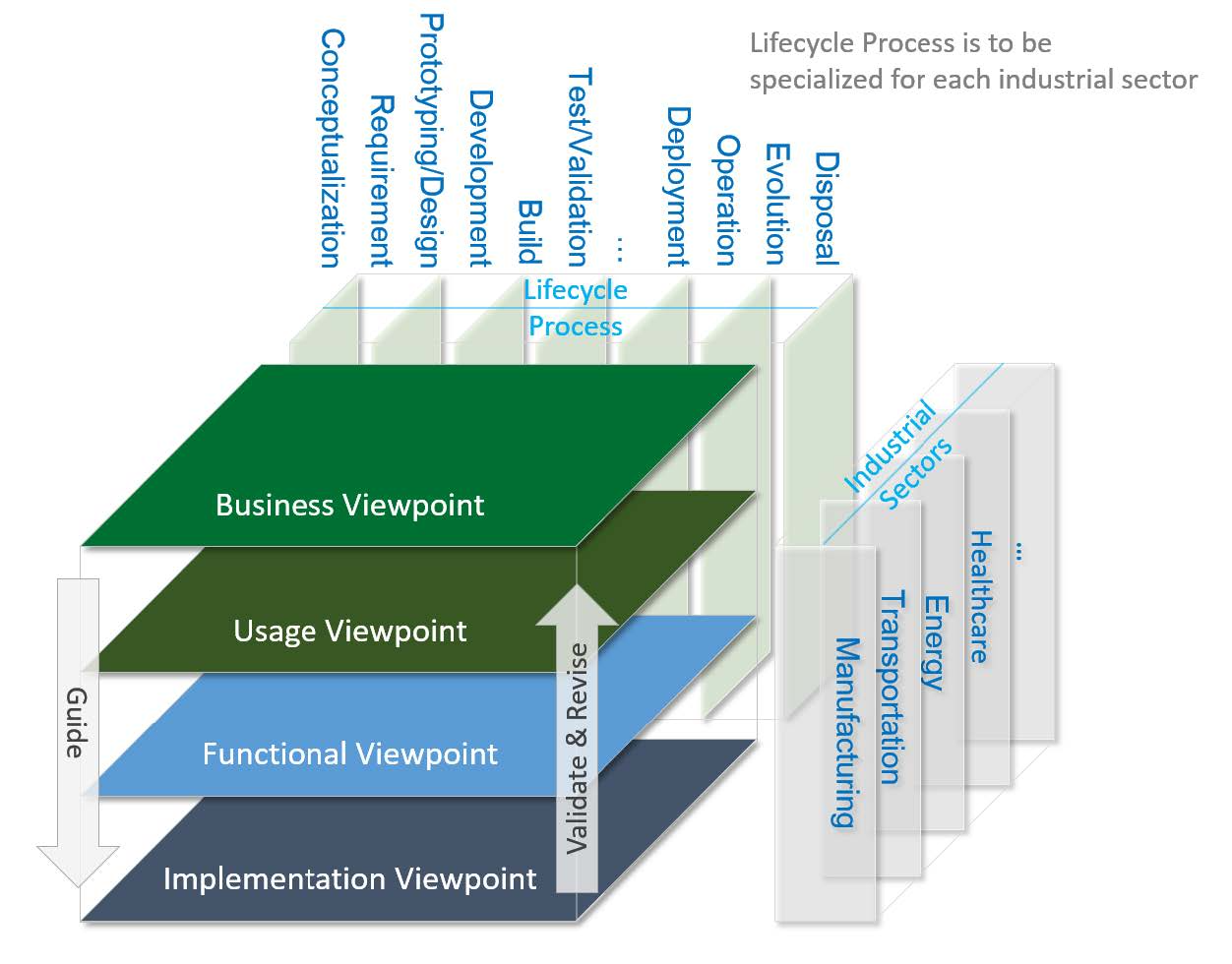}
  \caption{Industrial Internet Reference Architecture.\,\cite{Lin2017}}
  \label{fig:iira}
\end{figure}
Figure~\ref{fig:iira} depicts its three dimensions, comparable to the RAMI4.0. 
\begin{itemize}
	\item Viewpoints: As a result of analyzing various IIoT use cases, e.g. developed by the IIC, relevant stakeholders and their concerns are defined and associated to four viewpoints (business, usage, functional and implementation)
	\item Lifecycle Process: The IIRA is not a description of a system lifecycle process, which varies from one industrial sector to another. It is a tool for system conceptualization that highlights important system concerns that may affect lifecycle process. Through its viewpoints it provides guidance to system lifecycle processes from IIoT system conception, to design and implementation.
	\item Industrial Sectors: The related stakeholders in each viewpoints must regard the affected industrial sectors, to gather and describe the specific concerns resulting in the unique system requirements.
\end{itemize}
The IIRA summarizes common concerns from different perspectives of the stakeholders gathered from lots of use cases and projects, and therefore represents an high level of abstraction.
A central idea of the IIRA is the need to network larger systems and establish control over hierarchies of machines. 
Therefore this architecture seems well suited for Industrial Control Systems (ICS).
Accordingly, typical IIoT systems are decomposed into five functional, interconnected, domains (control, operations, information, application, business).  
The IIRA is meant as an iterative top-down process to describe and develop architectural concerns on each viewpoint by their stakeholders. 
The results of one viewpoint serve as a guideline and impose requirements on the viewpoint below, while discussing the concerns in a subsequent viewpoint may validate or cause a revision of the decisions in the viewpoint(s) above.
System architects may use and extend the architectural results of the implementation viewpoint as a basis for a concrete system architecture. 
For this purpose the IIRA describes some suitable architecture patterns, mainly based on the three-tier architecture pattern.

\subsection{5C-Architecture}
The 5C architecture is introduced by Lee et al. in \cite{Lee:2015} and focuses on Industrie 4.0 based manufacturing systems.
5C stands for the five levels: connection, conversion, cyber, cognition and configuration.
The architecture should be a guideline how to reach the goal of cognition (here called self-x) starting by the initial data acquisition.
So, this architecture aims to provide cognitive functions to the CPS.
\begin{figure}[htb]
  \centering
  \includegraphics[width=0.5\textwidth]{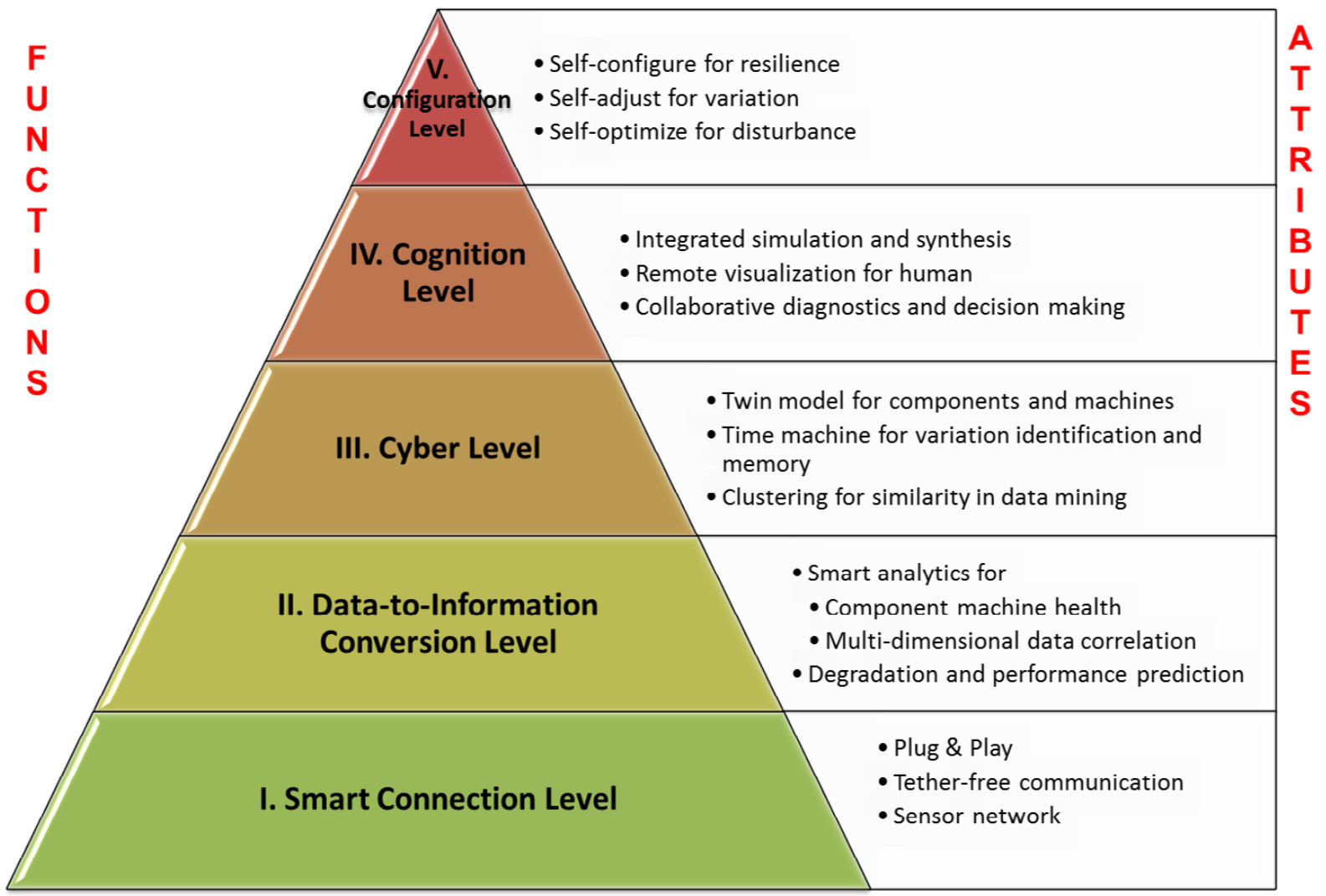}
  \caption{5C architecture for implementation of Cyber-Physical System.\,\cite{Lee:2015}}
  \label{fig:5C}
\end{figure}
The architecture is presented in Fig.\,\ref{fig:5C}.
On \textit{smart connection level}, data from versatile sources, such as sensors, controllers, MES, or ERP, are acquired by a central server.
Moreover, features, such as sensor signals, are selected on this level.
The \textit{data-to-information level} uses algorithms to process the data and generate information, such as health values or remaining useful life.
So typically machine learning techniques are implemented on this level.
The information from many machines are gathered in the \textit{cyber level}, which acts as central information hub.
It can be used to compare machines, make predictions, or get more insights of a machine through a combination of information or by using historical information.
The \textit{cognition level} provides deeper insights, especially to the user, which gets comparative information as well as information about single machines.
In the initial publication \cite{Lee:2015} it seems to be a human-machine-interface, whereas in later publication such as \cite{Lee:2017} the decision making is in the focus of this layer, which makes actually sense.
It can be used e.g. to optimize the maintenance by prioritizing the tasks, or logistic planning\,\cite{Lee:2017}.
\textit{Configuration level} is the most upper level, which provides feedback from the cyber space to the physical space and is a supervisory control that is needed for self-configuration and self-adaption.
In particular, it can be used for optimization or to increase the product quality.

\subsection{Soar}
The development of Soar was motivated by exploring the requirements for general intelligence, based on the human cognition theory.
The first version was created in 1982 as Soar 1 by John E. Laird, Allen Newell and Paul S. Rosenbloom, but to roots of Soar go back until 1956\,\cite{Laird:1987}.
The cognition theory evolved over time until current version Soar 9, which is available for free and still an active field of research\,\cite{Laird:2018}. 
Soar became a complex architecture since many new features have been added. This introduction focuses only on the most basic components and the basic functions.
\begin{figure}[htb]
  \centering
  \includegraphics[width=0.4\textwidth]{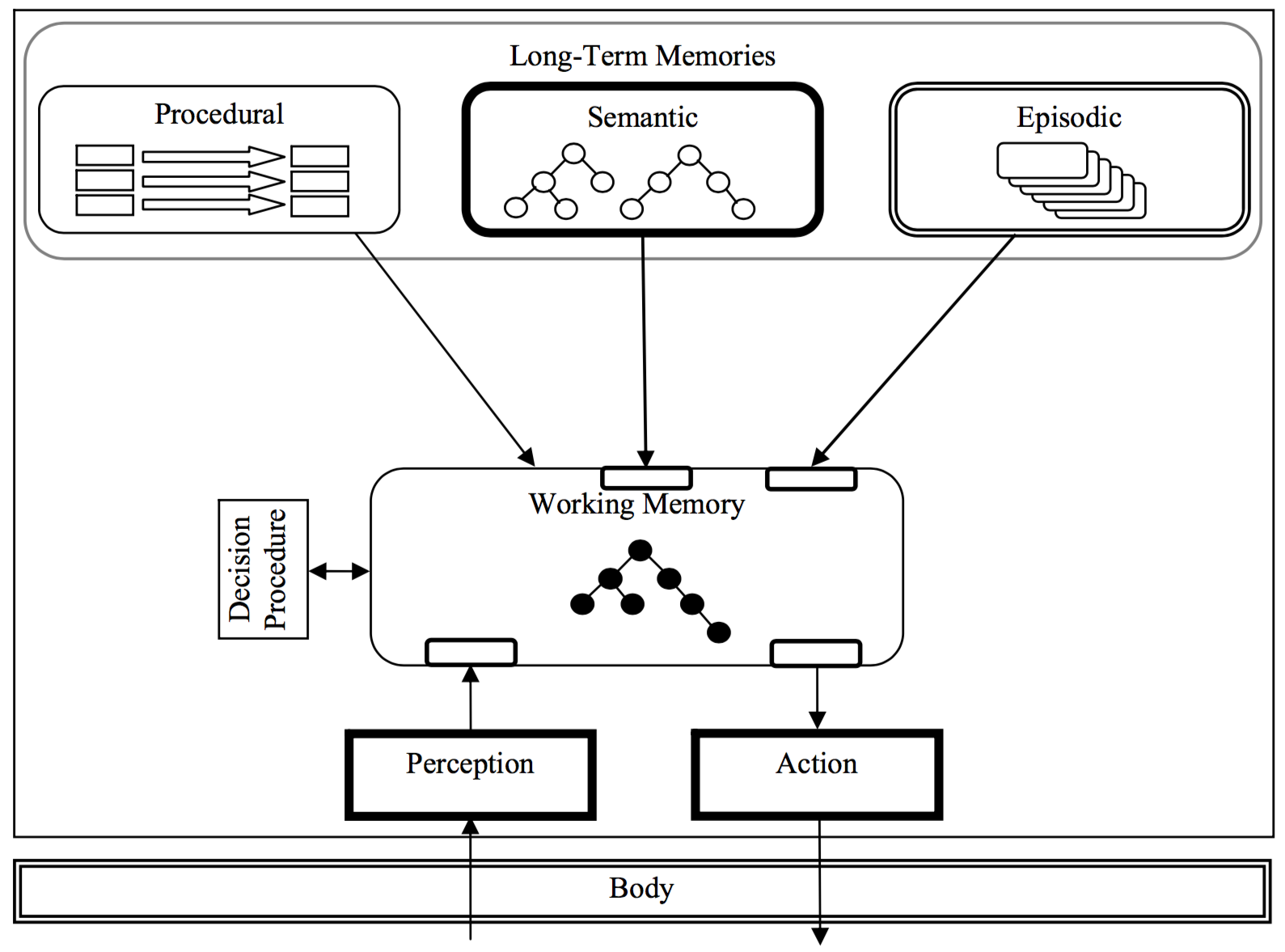}
  \caption{Simplified structure with focus on memories of Soar 9.\,\cite{Lehman:1996}}
  \label{fig:SOAR}
\end{figure}
The memory of Soar is divided into a long-term memory and a working memory, see Fig.\,\ref{fig:SOAR}. 
The working memory consists of objects that represent states, whereas the long-term memory consists of a procedural memory, semantic memory and episodic memory.
To perform a task, there is an initial state, a destination state, a problem space and operators that can be applied to change the states.
Production rules from the procedural memory are used to suggest an operator with certain preferences that can be applied.
A distinct operator is applied to the state. 
This sequence is repeated until the destination state has been reached.
If no distinct operator could be identified (called impasse), a subgoal is created, which is solved by trial and error.
Impasses are caused by a lack of knowledge.
If an impasse is solved once, chunking is used to preserve the knowledge by creating new production rules, to not run into the same impasse again.
\subsection{ACT-R}
In the Adaptive-Control-of-Thought-Rational (ACT-R) theory~\cite{Anderson:1996}, cognition is the result of an interaction of procedural and declarative knowledge. 
Procedural knowledge is represented in units called production rules, which mainly consists of a goal, actions and conditions. 
Declarative knowledge is represented in terms of chunks. 
The high level of connection between both becomes clear, as the state as well as the actions of production rules are stored as declarative knowledge.
Anderson's fundamental idea behind this theory is, that to reach cognition and intelligence: "The whole is no more than the sum of its parts, but it has a lot of parts"\,\cite{Anderson:1996}.
Anderson states, that intelligence is the result of gathering and tuning many small units of knowledge. 
The main questions in focus of the ACT-R research are the following:
\begin{itemize}
  \item How are these units of knowledge represented?
  \item How are they acquired?
  \item How are they deployed in cognition?
\end{itemize}
The visual and aural interfaces to the external world are responsible to create declarative knowledge chunks by appropriate encoding of the environment.
Manual and vocal functionality connected to the external world provide the ability to perform actions to the environment, e.g., steering moves in a car driving situation or using a computer keyboard to enter a solution for a math equation. 
The potentially large amount of knowledge gathered during the application of and required for cognition is only partly activated when a task occurs, according to its context and its prior success. 
This allows efficient access even if the amount of data gathered is quite large.

\section{Evaluation} \label{sec:evaluation}
To evaluate the existing architectures, we derived requirements from the use cases that are described in section\,\ref{sec:use-case}.
Additionally, we identified some general requirements, which cannot be derived from a single use case.
We merged similar requirements an present the overall requirements in Table\,\ref{tab:OverallRequ}.
The table represents a running number, the source of the requirements, i.e. where they are derived from, and a short description.
In this section, we compare the architectures scopes and the overall requirements.
After a broad overview, we describe the evaluation results for all architectures in detail.
Finally, we make a conclusion and identify the gap for cognitive architectures in CPPS.
\begin{table}[b]
\caption{Consolidation of requirements derived in section~\ref{sec:use-case} for a cognitive architecture in automation.}
\centering 
\begin{tabularx}{\columnwidth}{c|c|X} %
Result & Source & Description \\
\hline
R.1 & R-D.1 & Receive declarative goals from the user  \\
R.2 & R-D.2 & Specified interfaces are well defined  \\
R.3 & R-D.3 & Strategies to select a suitable algorithm \\
R.4 & R-D.4 & The system learns from experiences \\
R.5 & R-D.5 & Thorough knowledge representation \\
R.6 & R-A.1, R-B.1, R-C.1& Acquire data from distributed system \\
R.7 & R-A.2, R-B.2, R-C.2 & Store and manage acquired process data and models \\
R.8 & R-A.3, R-B.3, R-C.1 & Perform preprocessing   \\
R.9 & R-A.4, R-B.4, R-C.3 & Learn a model from data, might be time and resource limited \\
R.10 & R-A.5, R-B.5, R-C.5 & Perform a model analysis which might have a limited response time \\
R.11 & R-B.6 & Interaction with the user (HMI)\\ 
R.12 & R-A.6, R-C.4 & Decision making, e.g. sent new parameters to the controller\\
R.13 & R-A.7, R-C.6 & Apply the action on the controller \\
\end{tabularx}
\label{tab:OverallRequ} 
\end{table}
\subsection{Overall Evaluation}
We represent the evaluation results in Table\,\ref{tab:OverallEvaluation}.
Every requirement is in a single row, whereas the five reviewed architectures are each in a column.
A '-' is used to indicate that the requirement is not addressed by the architecture, a 'O' means that it is not sufficiently addressed, e.g. there may be some additional effort needed or it is not completely fulfilled, and a '\checkmark' indicates that the architecture fulfills the requirement.
Please keep in mind that the evaluation is not absolute and might look different, if it is e.g., performed in different use cases or in a different context than CPPS.
First of all, no reviewed architecture fulfills all requirements.
The automation architectures fit well, even though several requirements are unfulfilled.
Especially, requirement R.2 is not properly fulfilled by any architecture.
Just RAMI4.0 and IIRA cover this issue very roughly.
Requirements R.3 is also only partially fulfilled by two architectures, but this depends on the inserted model, so this issue can be satisfied.
Another interesting point is that the automation architectures fulfill similar requirements, whereas the cognitive architectures fulfill similar requirements, but those are completely different from the automation architectures.
There is obviously a gap of cognitive aspects in automation architectures, but cognitive architectures as known today are not suitable to fill this gap.
The boundary between both classes of architectures is between R.5 and R.6, which is also the boundary between use case specific and the general requirements.
So, the automation architectures are suitable for the use case requirements.
But obviously, there are some more requirements to construct a cognitive architecture that are not addressed until now.
\begin{table} 
\caption{Evaluation results for the architectures. Requirement addressed: \checkmark, not addressed: - , not sufficiently addressed: O}
\centering 
 \begin{tabular}{c|c|c|c|c|c} 
Requirement 	& RAMI4.0 	& IIRA 		& 5C 		& SOAR 		& ACT-R \\
\hline
R.1 			& - 			& - 				& - 			& \checkmark & \checkmark \\
R.2 			& O 			& O 				& - 			& - 			& - \\
R.3 			& - 			& - 				& - 			& O 			& O   \\
R.4 			& - 			& - 				& - 			& \checkmark & \checkmark   \\
R.5 			& \checkmark	& - 				& - 			& \checkmark & \checkmark   \\
R.6 			& \checkmark	& \checkmark   	& \checkmark	& - 			& -   \\
R.7 			& \checkmark	& \checkmark		& O			& -			&  -    \\
R.8 			& \checkmark	& \checkmark		& O 			& -			&   -   \\
R.9 			& O			& \checkmark 	& \checkmark & -  		& -   \\
R.10			& O			& \checkmark 	& \checkmark & - 		& -    \\
R.11  		& \checkmark & O				& - 	  		& - 			& -    \\
R.12 	    & O   		&  \checkmark  	& \checkmark & \checkmark & \checkmark \\
R.13 		& -			&  \checkmark	& \checkmark & \checkmark & \checkmark \\
\end{tabular}
\label{tab:OverallEvaluation}
\end{table}
\subsection{Reference Architecture Model Industrie 4.0 (RAMI4.0)}
All requirements that were derived from the use cases are implemented on the Layers-axis of RAMI4.0 model. The following descriptions concerning the different layers stem from \cite{VDI:2015}. Data can be exchanged between different functions within a layer and all adjacent layers. The \textit{Communication Layer} is meant to "unify communication by using a consistent data format towards the Information Layer". The \textit{Information Layer} then "serves structured data via interfaces". There is no other description or definition for other interfaces between layers (R.2). It is also not possible for the architecture to receive declarative goals from the user (R.1) or for the system to use a learned strategy to select the best algorithm for a problem (R.3) as the system does not learn from experiences in a cognitive sense (R.4). Knowledge representation (R.5) is located in the \textit{Business Layer}. There the business models, legal and regulatory conditions as well as general rules that the system has to obey are represented. The \textit{Integration Layer} provisions asset information and generates events (R.6). The \textit{Information Layer} persists data that represents the models (R.7). Pre-processing is also done by the \textit{Information Layer}, that works as runtime environment and accepts and transforms data for the \textit{Functional Layer} (R.8). The high grade of abstraction makes it hard to use the architecture as a guideline to implement a software system for the described use cases. Those require model learning and analysis, simulation with new data and decision making (R.9, R.10, R.12). The \textit{Functional Layer} contains "runtime and modelling environments for services that support business processes", but remains unclear about implementation of actual functionality. The \textit{Information Layer} implements "interaction with the user via human-machine-interface" (R.11). RAMI4.0 does not intend to lead derived actions back to the controller (R.13). Temporary remote access is allowed for maintenance purposes but not for regular functional integration. It is also specified that it is sufficient for a I4.0 component to possess passive communication abilities, i.e, providing their information to another system, maybe even through another system. Therefore RAMI4.0 unifies the perspectives of various fields in the different dimensions. This may be simultaneously the biggest advantage and disadvantage. For mechanical and electrical engineering virtually any asset can be classified which aids design of I4.0 components. The additional work \cite{ZVEI:2016} \cite{Bmwi:2018} specifies which data needs to be stored and tries to standardize communication and access to functionality. But for the design and implementation of a CPPS software system the architecture may be too abstract. 

\subsection{Industrial Internet Reference Architecture (IIRA)}
The IIRA is developed as a template and methodology that enables and guides system architects to design their IIoT systems based on a common framework. 
That said, it becomes clear, that the IIRA does not address all requirements, neither all requirements from the use cases, nor from general aspects, as shown in Table\,\ref{tab:OverallEvaluation}.
Declarative goals (R.1) are not explicitly addressed by the IIRA, and accurate interface definitions (R.2) come into place by further development of the results from the IIRA by a system architect.  
Cognitive aspects are not explicitly addressed as well (R.3, R.4). 
Expert knowledge seems to be a responsibility of the business domain of the functional viewpoint, but no further specification of the knowledge representation is given.
The use case specific requirements (R.6-R.13) are in focus of the IIRA. 
Ingesting sensor data and data quality processing (R.6-R.8) as well as storage and distribution of the data, modelling the state of the IIoT system and performing simulation and optimization tasks (R.9, R.10) are located in the information domain in the functional viewpoint of the architecture.
Human machine interfaces (R.11) are located in the business domain in the functional viewpoint. 
The control domain is located in the functional viewpoint as well, and addresses the connection of sensors, controllers, actuators, gateways and other edge systems (R.12, R.13).

\subsection{5C-Architecture}
Even if other architectures fulfill more requirements than the 5C architecture does, it stands out with a low abstraction and it aims to similar use cases.
The 5C architecture is able to acquire data, process them, learn from experiences and adapt the machines, see Table\,\ref{tab:OverallEvaluation}.
But some requirements are not or not properly fulfilled.
One main issue is the missing HMI of the 5C, so it neither can capture declarative goals from the user (R.1) nor inform the user (R.11).
The other main issue is that interfaces are not defined properly (R.2), there is no explanation of any interface or the transmitted data. 
The cognitive aspect, to learn from experiences (R.4), is limited to case-based reasoning, where the current situation is compared to similar situations, e.g. to estimate time to failures, which does not address R.4.
The selection of a suitable algorithm (R.3) is another cognitive aspect with is not addressed in the 5C architecture.
Some minor aspects are that it is not possible to integrate expert knowledge (R.5), data are just stored after computing, which leads to a loss of process data, and the preprocessing is not defined as a separate step, but since it can be connected to an algorithm it is indicated with a O in Table\,\ref{tab:OverallEvaluation}.
\subsection{Soar and ACT-R}
First of all it is necessary to clarify that, due to the similarity of both architectures, ACT-R and Soar, are evaluated together. 
A fundamental feature is the ability to state goals in a declarative manner (R.1). 
Although the architectures define their interfaces thoroughly, the modular extendability (R-2, see Section\,\ref{sec:use-case}) is not addressed by the cognitive architectures.
Strategies for algorithm selection (R-3) are not directly implemented, but as this could be implemented as a cognitive task and learned by the system, we evaluate this requirement as fulfillable with workarounds. 
(R-4) can be seen as fulfilled, as this is the major task of both architectures. 
Most requirements from the use cases are not explicitly in focus of the architecture (R-6 to R-11). 
However, decision making, as a result of cognition, and sending parameters to the controller and apply actions (R-12, R-13) is implemented by both architectures.
\subsection{Conclusion}
\begin{figure}[htb]
  \centering
  \includegraphics[width=0.3\textwidth]{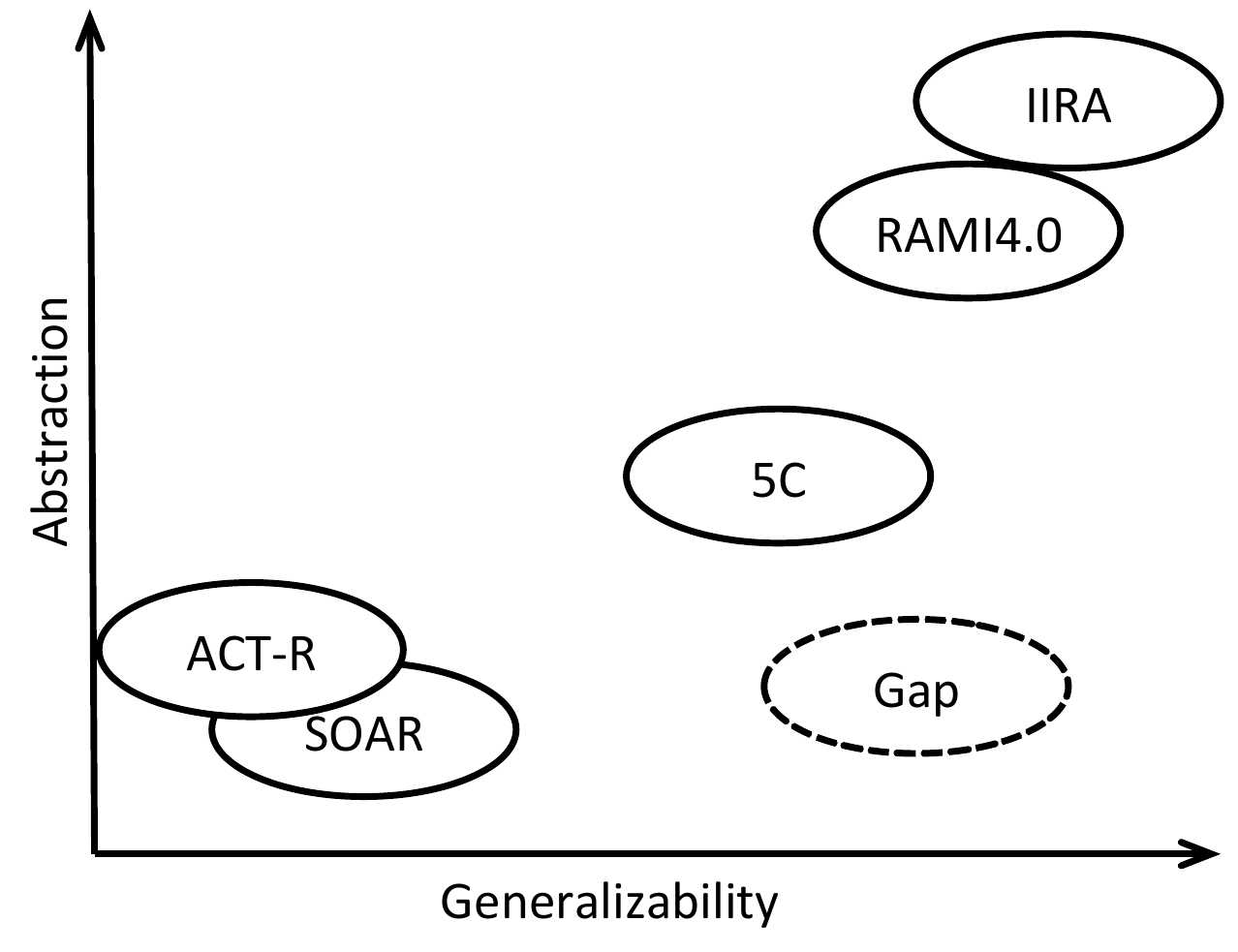}
  \caption{Visualization of the Gap in current architectures}
  \label{fig:Conclusion}
\end{figure}

None of the introduced architectures fulfills all requirements of a cognitive architecture, see Table~\ref{tab:OverallRequ}.
In Fig.\,\ref{fig:Conclusion} we classified the architectures regarding their level of abstraction and their generalizability. 
A high level of abstraction means that the implementation has no boundaries, so there is a lot of freedom in implementation, whereas a high generalizability means that the architecture can be adapted to many different use cases.
The cognitive architectures have a low abstraction, because there are fixed structures to implement a use case.
Although the architecture is cognitive, it has to be adapted for different use cases, since all use cases require different knowledge.
In comparison to that, the automation architectures are abstract.
There are no strong limitation for their implementation, which makes it challenging.
But therefore, they are generic, so they could be applied to many use cases.
The 5C architecture is less generic and less abstract than the other automation architectures, but it is still too abstract for an easy usage for the use cases.
The gap is an architecture with low abstraction, so that it is easy to implement, and a high generalizability, to fit to many CPPS use cases.
This can be achieved by combining the two types of architectures. 
Additionally, the interfaces have to be well defined, to enable modularity and the reuse of existing parts.
This need is also mentioned for RAMI4.0\,\cite{VDI:2015} and for the IIRA~\cite{Lin2017}, where the interfaces are defined in the implementation viewpoint.
The 5C architecture is a good starting point, but there is the need to make it more concrete.

\section{Summary and Outlook}
This paper focuses on the review and evaluation of reference architectures for the implementation of CPPS in I4.0 environment with additional cognitive abilities. Therefore reference architectures from the field of automation as well as architectures that stem from the field of cognitive sciences are considered.
The future goal is to develop a reference architecture suitable to be employed in several real-world use cases, three of these are introduced in this work. 
Altogether they require several well known tasks in I4.0 systems, i.e., condition monitoring, predictive maintenance, root cause analysis and optimization. 
In a first step, requirements that this architecture has to fulfil in the field of CPPS, are analyzed (C1).
The results reveal that on the one hand there are general requirements, which are not use case specific, and requirements which are use case specific. 
Some of the requirements arising from the use cases may be aggregated, as there are several use cases with exactly this or similar requirements.
The second step is the evaluation of the considered architectures according to the requirements (C2). 
Two classes of architectures come into focus, reference architectures from an automation perspective, and cognitive architectures from the field of cognitive science.
Expectably, the reference architectures fit well into the use case based requirements, and the cognitive architectures are suitable to enable a high level of adaptability.
However, none of the regarded architectures fulfill all requirements, and additionally a gap between the two classes of the architectures could be uncovered (C3).
It becomes apparent that architectures with a low level of abstraction also are not suitable to address many use cases efficiently. 
On the other hand, architectures with an high abstraction degree, suitable for many use cases, are not easy to implement nor easy to be specified in a necessary level of detail. 
Future work will focus on the necessary tasks to fill the uncovered gap. 
Two classes of architectures with different degrees of abstraction from different fields of science (engineering and cognitive science) shall be combined smartly, to reach following goals:
\begin{enumerate}
\item[a)] Easy implementation and expandability by thorough modularity.
\item[b)] Efficient usage for several use cases.
\end{enumerate}
It seems that the 5C architecture comes closest to this goals so far. The next step is the further investigation of this architecture to reveal the necessary tasks of enhancements of the 5C to fill the gap towards increasing generalizability and decreasing abstraction. 

\section*{Acknowledgment}
The work was supported by the German Federal Ministry of Education and Research (BMBF) under the project "KOARCH" (funding code: 13FH007IA6 and 13FH007IB6).

\bibliographystyle{IEEEtran}

\bibliography{%
literature%
}

\begin{thebibliography}{10}
\providecommand{\url}[1]{#1}
\csname url@samestyle\endcsname
\providecommand{\newblock}{\relax}
\providecommand{\bibinfo}[2]{#2}
\providecommand{\BIBentrySTDinterwordspacing}{\spaceskip=0pt\relax}
\providecommand{\BIBentryALTinterwordstretchfactor}{4}
\providecommand{\BIBentryALTinterwordspacing}{\spaceskip=\fontdimen2\font plus
\BIBentryALTinterwordstretchfactor\fontdimen3\font minus
  \fontdimen4\font\relax}
\providecommand{\BIBforeignlanguage}[2]{{%
\expandafter\ifx\csname l@#1\endcsname\relax
\typeout{** WARNING: IEEEtran.bst: No hyphenation pattern has been}%
\typeout{** loaded for the language `#1'. Using the pattern for}%
\typeout{** the default language instead.}%
\else
\language=\csname l@#1\endcsname
\fi
#2}}
\providecommand{\BIBdecl}{\relax}
\BIBdecl

\bibitem{Lee2008}
E.~A. Lee, ``{Cyber physical systems: Design challenges},'' \emph{Proceedings -
  11th IEEE Symposium on Object/Component/Service-Oriented Real-Time
  Distributed Computing, ISORC 2008}, pp. 363--369, 2008.

\bibitem{Bangemann2015}
T.~Bangemann \emph{et~al.}, ``{Industrie 4.0 – Technical Assets; Grundlegende
  Begriffe, Konzepte, Lebenszyklen und Verwaltung},''
  \emph{VDI/VDE-Gesellschaft Mess- und Automatisierungstechnik (GMA)}, Nov
  2015.

\bibitem{Bitkom:2015}
\BIBentryALTinterwordspacing
Bitkom, VDMA, and ZVEI, ``{Umsetzungsstrategie Industrie 4.0. - Ergebnisbericht
  der Plattform Industrie 4.0},'' \emph{Plattform Industrie 4.0}, p. 100, Apr
  2015. [Online]. Available:
  \url{https://www.bitkom.org/sites/default/files/file/import/150410-Umsetzungsstrategie-0.pdf}
\BIBentrySTDinterwordspacing

\bibitem{Deloitte2018}
\BIBentryALTinterwordspacing
Deloitte, ``{The fourth industrial revolution is here: Are you ready?}''
  \emph{Deloitte Insight}, pp. 1--26, 2018. [Online]. Available:
  \url{https://www2.deloitte.com/content/dam/Deloitte/de/Documents/Innovation/Industry-4-0-Are-you-ready-report.pdf}
\BIBentrySTDinterwordspacing

\bibitem{Bunte:2019}
A.~Bunte, B.~Stein, and O.~Niggemann, ``Model-based diagnosis for
  cyber-physical production systems based on machine learning and
  residual-based diagnosis models.''\hskip 1em plus 0.5em minus 0.4em\relax
  Hawaii, USA: Thirty-Third AAAI Conference on Artificial Intelligence
  (AAAI-19), Jul 2019.

\bibitem{VBS:2017}
K.~Bauer \emph{et~al.}, ``Benefits of application scenario value-based
  service,'' Federal Ministry for Economic Affairs and Energy (BMWi), Tech.
  Rep., Apr. 2017.

\bibitem{Maier:2015}
A.~Maier and O.~Niggemann, ``On the learning of timing behavior for anomaly
  detection in cyber-physical production systems,'' in \emph{International
  Workshop on the Principles of Diagnosis (DX)}, Aug 2015.

\bibitem{Jodlbauer:2016}
H.~Jodlbauer and M.~Schagerl, ``Reifegradmodell industrie 4.0 - ein
  vorgehensmodell zur identifikation von industrie 4.0 potentialen,'' in
  \emph{Informatik 2016}, H.~C. Mayr and M.~Pinzger, Eds.\hskip 1em plus 0.5em
  minus 0.4em\relax Bonn: Gesellschaft f\"ur Informatik e.V., 2016, pp.
  1473--1487.

\bibitem{Lehman:1996}
J.~F. Lehman, J.~Laird, and P.~S. Rosenbloom, ``A gentle introduction to soar ,
  an architecture for human cognition : 2006 update,'' 1996.

\bibitem{VDI:2015}
\BIBentryALTinterwordspacing
P.~Adolphs \emph{et~al.}, ``{Referenzarchitekturmodell Industrie 4.0
  (RAMI4.0)},'' \emph{VDI /VDE Statusreport}, p.~32, Apr 2015. [Online].
  Available:
  \url{https://www.vdi.de/fileadmin/user_upload/VDI-GMA_Statusreport_Referenzarchitekturmodell-Industrie40.pdf}
\BIBentrySTDinterwordspacing

\bibitem{SGAM:2012}
\BIBentryALTinterwordspacing
{CEN-CENELEC-ETSI Smart Grid Coordination Group}, ``{Smart Grid Information
  Security},'' pp. 1--107, Nov 2012. [Online]. Available:
  \url{ftp://ftp.cen.eu/EN/EuropeanStandardization/HotTopics/SmartGrids/Security.pdf}
\BIBentrySTDinterwordspacing

\bibitem{ZVEI:2016}
\BIBentryALTinterwordspacing
{ZVEI}, ``{Beispiele zur Verwaltungsschale - Basisteil},'' 2016. [Online].
  Available:
  \url{https://www.zvei.org/presse-medien/publikationen/beispiele-zur-verwaltungsschale-der-industrie-40-komponente-basisteil/}
\BIBentrySTDinterwordspacing

\bibitem{Lin2017}
S.-W. Lin \emph{et~al.}, ``{The Industrial Internet of Things Volume G1:
  Reference Architecture v1.80},'' Industrial Internet Consortium, Tech. Rep.,
  Nov 2017.

\bibitem{Lee:2015}
\BIBentryALTinterwordspacing
J.~Lee, B.~Bagheri, and H.-A. Kao, ``A cyber-physical systems architecture for
  industry 4.0-based manufacturing systems,'' \emph{Manufacturing Letters},
  vol.~3, pp. 18 -- 23, 2015. [Online]. Available:
  \url{http://www.sciencedirect.com/science/article/pii/S221384631400025X}
\BIBentrySTDinterwordspacing

\bibitem{Lee:2017}
\BIBentryALTinterwordspacing
J.~Lee, C.~Jin, and B.~Bagheri, ``Cyber physical systems for predictive
  production systems,'' \emph{Production Engineering}, vol.~11, no.~2, pp.
  155--165, Apr 2017. [Online]. Available:
  \url{https://doi.org/10.1007/s11740-017-0729-4}
\BIBentrySTDinterwordspacing

\bibitem{Laird:1987}
\BIBentryALTinterwordspacing
J.~E. Laird, A.~Newell, and P.~S. Rosenbloom, ``Soar: An architecture for
  general intelligence,'' \emph{Artif. Intell.}, vol.~33, no.~1, pp. 1--64,
  Sep. 1987. [Online]. Available:
  \url{http://dx.doi.org/10.1016/0004-3702(87)90050-6}
\BIBentrySTDinterwordspacing

\bibitem{Laird:2018}
\BIBentryALTinterwordspacing
J.~Laird and S.~Mohan, ``Learning fast and slow: Levels of learning in general
  autonomous intelligent agents,'' in \emph{AAAI Conference on Artificial
  Intelligence}, 2018. [Online]. Available:
  \url{https://www.aaai.org/ocs/index.php/AAAI/AAAI18/paper/view/17261}
\BIBentrySTDinterwordspacing

\bibitem{Anderson:1996}
J.~R. Anderson, ``{A Simple Theory of Complex Cognition},'' \emph{American
  Psychologist}, 1996.

\bibitem{Bmwi:2018}
\BIBentryALTinterwordspacing
{Plattform Industrie 4.0}, ``{Welche Kriterien m{\"{u}}ssen Industrie-4.0-
  Produkte erf{\"{u}}llen? Leitfaden 2018},'' 2018. [Online]. Available:
  \url{https://www.de.digital/DIGITAL/Redaktion/DE/Publikation/leitfaden-industrie-40-produkte.pdf?__blob=publicationFile&v=6}
\BIBentrySTDinterwordspacing

\end{thebibliography}

\end{document}